\documentclass{JHEP3}
\usepackage[dvips]{graphicx}
\usepackage{amsmath,amssymb}

\newcommand{\be}{\begin{equation}}
\newcommand{\ee}{\end{equation}}
\newcommand{\bea}{\begin{eqnarray}}
\newcommand{\eea}{\end{eqnarray}}
\newcommand{\fatk}{\mathbf k}

\newcommand{\fate}{\mathbf e}
\newcommand{\mat}{\left ( \begin{array}{cc}}
\newcommand{\emat}{\end{array} \right )}

\newcommand{\bfa}{{\mathbf a}}

\newcommand{\bfe}{{\mathbf e}}
\newcommand{\bfk}{{\mathbf k}}
\newcommand{\bfl}{{\mathbf l}}

\newcommand{\bfq}{{\mathbf q}}

\newcommand{\bz}{\bar{z}}

\newcommand{\bpsi}{\bar{\psi}}

\newcommand{\bxi}{\bar{\xi}}

\newcommand{\bsigma}{\bar{\sigma}}

\newcommand{\hQ}{\widehat{Q}}
\newcommand{\hkappa}{\hat{\kappa}}

\newcommand{\cN}{{\cal N}}

\newcommand{\Z}{\mathbb{Z}}

\newcommand{\nn}{\nonumber}

\newcommand{\Tr}{{\rm Tr}\,}

\newcommand{\e}{\epsilon}

\title{Lattice Supersymmetry: Equivalence between the 
Link Approach and Orbifolding
}
\author{Poul H. Damgaard%
\footnote{phdamg@nbi.dk}  
and So Matsuura%
\footnote{matsuura@nbi.dk}\\ 
The Niels Bohr Institute, 
The Niels Bohr International Academy,
Blegdamsvej 17, DK-2100 Copenhagen, Denmark}
\abstract{We examine the relation between 
supersymmetric lattice gauge
theories constructed by the link approach and by orbifolding 
and show that they are equivalent. We discuss the number
of preserved supersymmetries.}

\begin{document}

\section{Introduction}

A number of apparently different ways to preserve exactly
both gauge symmetry and some supersymmetries on a euclidean space-time lattice 
have been proposed. One is based on the so-called orbifolding method 
\cite{Kaplan:2002wv}--\nocite{Cohen:2003xe}%
\nocite{Cohen:2003qw}\nocite{Kaplan:2005ta}\cite{Damgaard:2007be}, 
in which dimensional reduction to a zero-dimensional
mother theory is followed by an orbifold projection that leaves
invariant both a fixed number of supersymmetries and a discrete
symmetry. The latter can be viewed as translations on a 
space-time lattice that appears as a
result of ``deconstruction''. Two apparently different formulations,
pursued by Catterall
\cite{Catterall:2003wd}--%
\nocite{Catterall:2004np}\cite{Catterall:2005fd}
and Sugino \cite{Sugino:2003yb}--\nocite{Sugino:2004qd}%
\cite{Sugino:2004uv}\nocite{Sugino:2006uf},
use as
starting points ideas from topological field theory in order
to preserve exactly
a number of supersymmetries on a space-time lattice. Recently it has
been demonstrated how both of these formulations
can be understood from the point of view of
orbifolding and deconstruction as well 
\cite{Takimi:2007nn}\cite{Damgaard:2007xi}. 
Finally, an
approach that is again tied up closely with both topological
field theory (twisted supersymmetry) and the Dirac-K\"{a}hler
formulation of lattice fermions has been advocated 
\cite{D'Adda:2004jb}--\nocite{D'Adda:2005zk}\cite{D'Adda:2007ax}. 
We will call this latter approach to lattice supersymmetry
for the {\em link approach}. What is unique to that formulation is
the claim that it can preserve exactly {\em all} supersymmetries at
finite lattice spacings, not just those associated with the nilpotent
charges related to the underlying topological field theories. There
has recently been some discussion about this issue 
\cite{Bruckmann:2006ub}\cite{Bruckmann:2006kb}. 

Because also the result of the link approach resembles so much that
of orbifolding, one would like to understand better the relationship 
between the two formalisms. In this paper we show that the 
link approach is completely equivalent to the one based
orbifolding. We limit ourselves to describing this equivalence
in detail for the case of two-dimensional ${\cal N} = (2,2)$ 
supersymmetric Yang-Mills theory. It will be clear from our
discussion that the equivalence trivially generalizes. 
This equivalence
between the two formulations makes it more urgent to understand
also the number of preserved supersymmetries on the lattice.
This prompts us to investigate the fate of those supersymmetries
that are lost in the orbifolding procedure, and only hoped to
be regained in the continuum limit. As we shall show, the
additional supersymmetry transformations of the link approach 
have a natural explanation in terms of the orbifolding
procedure. As expected, they correspond to field transformations
that violate the Leibniz rule of field variations, and 
we thus cannot see these additional transformations as symmetries
of the action. Nevertheless, the origin of these transformations
can be clearly understood from the orbifolding point of view.
In this way, the apparent discrepancy in terms of
the number of preserved supersymmetries in
the two formulations is resolved.

The organization of this paper is as follows. 
In the next section, we briefly review the orbifold projection of 
the zero-dimensional Yang-Mills
matrix theory (the mother theory), and explain how several of
the supersymmetries are
broken by the orbifold projection. We compare the action with
that of the link approach, and show how the most general
orbifolded action (which does not preserve any supercharges
at all) is in one-to-one correspondence with that of the
link approach. The shift parameters of the link approach are
identified with the $U(1)$ charges of the supersymmetry generators
in the orbifolded action.
In section 3, we investigate the fate of the broken supersymmetries.
We show that the would-be
transformations agree exactly with those of the link approach 
if we allow for a redefinition of the fermionic parameters,
and we discuss the interpretation of these supersymmetry transformations.
Section 4 contains our conclusions.

\section{Supersymmetry transformations in the mother theory}

We begin by briefly recalling the main ingredients in the orbifold
construction of supersymmetric lattice gauge theories.%
\footnote{For a
nice review, see, $e.g.$, ref. \cite{Giedt:2006pd}. }
Because the points
we shall focus on are not specific in
regards to, for example, dimensionality, we restrict ourselves to
the ${\cal N} = (2,2)$ supersymmetric gauge theory in two
space-time dimensions. 
The action of the mother theory is in this case
obtained by dimensional reduction 
of four-dimensional $\cN=1$ supersymmetric Yang-Mills theory:
\begin{equation}
 S_m = \frac{1}{g^2}\Tr \Bigl(\frac{1}{4}v_{\alpha\beta}^2
+\bpsi \bsigma_\alpha [v_\alpha,\psi]\Bigr),
\label{mother action}
\end{equation}
where $\alpha,\beta=0,\cdots,3$, 
$v_\alpha$ are Hermitian bosonic 
matrices, $\psi$, $\bpsi$ are independent two-components spinors, 
and $v_{\alpha\beta}=i[v_\alpha,v_\beta]$. 
For the purpose for the future discussion, we assume the gauge group 
of the theory to be $U(kN^2)$. 
In the following, we use the notation, 
\begin{equation}
 \sigma_{\alpha}=(1_2,-i\tau_i), \qquad 
 \bsigma_{\alpha}=(1_2,i\tau_i), 
\end{equation}
where $\tau_i$ ($i=1,2,3$) are the Pauli matrices. 
In addition to the gauge symmetry, 
$v_\alpha \to g^{-1} v_\alpha g, \cdots$, 
this theory is invariant under
the ``global'' symmetry $SO(4)\times U(1)$, which corresponds to 
the Lorentz symmetry and the R-symmetry of the four-dimensional $\cN=1$
SYM theory \cite{Cohen:2003xe}, respectively. 
Furthermore, 
the action (\ref{mother action}) is invariant under the following
supersymmetry transformation: 
\begin{align}
 \delta v_\alpha &= -i \bpsi \bsigma_\alpha \xi + i \bxi
 \bsigma_\alpha \psi, \nn \\
 \label{supertrans mother}
 \delta \psi &= -i v_{\alpha\beta}\sigma_{\alpha\beta}\xi, \\
 \delta \bpsi &= i v_{\alpha\beta}\bxi \bsigma_{\alpha\beta}, \nn
\end{align}
where $\xi$ and $\bxi$ are constant Grassmann-odd spinor parameters. 

Following ref. \cite{Cohen:2003xe}, we define complex fields $z_m$ and
$\bz_m$ ($m=1,2$) by 
\begin{alignat}{2}
 z_1 &= -iv_1 + v_2, &\quad \bz_1 &= iv_1 + v_2, \nn \\
 \label{redefined fields boson}
 z_2 &= v_0 + iv_3, & \bz_2 &= v_0 -i v_3 , 
\end{alignat}
and express the component fields of $\psi$ and $\bpsi$ as 
\begin{equation}
 \psi = \left(\begin{matrix}\chi_{12} \\ \eta \end{matrix} \right), 
 \qquad 
 \bpsi =\left(\psi_1,\psi_2\right).
 \label{fermion components}
\end{equation}
Using these fields, 
the action of the mother theory (\ref{mother action}) can be rewritten
as 
\begin{align}
 S_m = \frac{1}{g^2}\Tr \Bigl(
\frac{1}{4}\left|[z_m,z_n]\right|^2 + \frac{1}{8}[z_m,\bz_m]^2
+ \eta[\bz_m,\psi_m] -\chi_{mn}[z_m,\psi_n]
\Bigr). 
\label{mother action 2}
\end{align}
In this expression, the global $U(1)$ symmetries are manifest. 
In fact, one can easily show that there are three independent $U(1)$
symmetries for which all the fields (\ref{redefined fields boson})
and (\ref{fermion components})  
have definite charges $q_a$ $(a=1,2,3)$ 
as shown in table \ref{charge table 4 SUSY}. 
\begin{table}[h]
\caption{The charge assignment of the maximal $U(1)$ symmetries} 
\begin{center}
\begin{tabular}{c|cccccc}
 & $z_1$ & $z_2$ & $\eta$ & $\chi_{12}$ & $\psi_1$ & $\psi_2$ \\
\hline
$q_1$ & 1 & 0 & 1/2 & -1/2 & 1/2 & -1/2 \\
$q_2$ & 0 & 1 & 1/2 & -1/2 & -1/2 & 1/2 \\
$q_3$ & 0 & 0 & 1/2 & 1/2 & -1/2 & -1/2 \\
\end{tabular}
\end{center}
\label{charge table 4 SUSY}
\end{table}
In terms of the new variables,
the supersymmetry transformations (\ref{supertrans mother}) take 
the forms
\begin{align}
 \delta z_m &= 2i \hkappa \psi_m + 2i \hkappa_m \eta, \nn \\
 \delta \bz_m &= -2i \hkappa_{mn}  \psi_n 
  -2i \hkappa_n \chi_{mn}, \nn \\
\label{supertrans mother 2}
\delta \eta &= \frac{i}{2}\hkappa [z_m,\bz_m] 
 +\frac{i}{2}
 \hkappa_{mn}[z_m,z_n], \\
\delta \chi_{12} &= -i\hkappa [\bz_1,\bz_2] 
 -\frac{i}{2}\hkappa_{12}[z_m,\bz_m], \nn \\
\delta \psi_m &= i\hkappa_n \left(
[z_m,\bz_n]-\frac{1}{2}\delta_{mn}[z_l,\bz_l]
\right), \nn
\end{align}
where we have expressed the components of $\xi$ and $\bxi$ as 
\begin{equation}
 \xi = \left(\begin{matrix}\hkappa_{12} \\ \hkappa\end{matrix}\right), 
\qquad 
 \bxi = \left(\hkappa_1, \hkappa_2\right), 
\end{equation} 
with $\hkappa_{mn}=-\hkappa_{nm}$.
We emphasize here that the transformations (\ref{supertrans mother 2}) 
correspond to a symmetry of the action if and only if the supersymmetry 
parameters 
$\hkappa$, $\hkappa_m$ and $\hkappa_{12}$ transform as singlets 
under the gauge group. This elementary fact, which is also obvious from
the transformation law (\ref{supertrans mother}), is crucial for the
discussion of the number of preserved supersymmetries below. 
We can define the operators of supercharges
$\{\hQ,\hQ_m,\hQ_{12}\}$ through the transformation (\ref{supertrans mother 2})
as 
\begin{equation}
 \delta \Phi = 2i \hkappa \hQ\Phi 
- 2i \hkappa_{12} \hQ_{12}\Phi + 2i \hkappa_m \hQ_m \Phi ~, 
 \label{definition of Q}
\end{equation}
where $\Phi$ is a generic field of the theory. 
It is straightforward to show that the supercharges satisfy the algebra, 
\begin{equation}
 \{\hQ,\hQ_m\} = -\frac{1}{2}[\bz_m, \,\cdot\,], \quad 
 \{\hQ_{12},\hQ_m\} = \frac{1}{2}\e_{mn}[z_n, \,\cdot\,],  
\label{algebra}
\end{equation}
with the other anticommutators vanishing, up to use of the equations of motion. 

Next, we carry out the orbifold projection. 
In order to obtain a two-dimensional lattice formulation, we follow the
standard procedure and mod out by $Z_N\times Z_N$ which is a 
subgroup of the full symmetry group of the mother theory 
\cite{Kaplan:2002wv}--\nocite{Cohen:2003xe}\nocite{Cohen:2003qw}%
\cite{Kaplan:2005ta}. 
In this projection, the $U(1)$ charges of the fields play crucial
roles. As mentioned above, the mother theory has three independent $U(1)$
symmetries and any linear combination of them is also a symmetry of the
theory. Following \cite{Damgaard:2007be}, we define two $U(1)$
charges so that $\eta$ has zero charges, 
\bea
r_1 &~\equiv~ & \ell_1^1q_1 + \ell_1^2q_2 - (\ell_1^1+\ell_1^2)q_3, \cr
r_2 &~\equiv~ & \ell_2^1q_1 + \ell_2^2q_2 - (\ell_2^1+\ell_2^2)q_3 . 
\label{integer U(1) charges 4 SUSY}
\eea
Introducing two vectors, 
\be
{\mathbf e}_1 ~\equiv~ {\ell_1^1 \choose \ell_2^1}, \quad
{\mathbf e}_2 ~\equiv~ {\ell_1^2 \choose \ell_2^2}, 
\ee
the charge assignments under these $U(1)$'s are given 
in Table \ref{remaining charge table 4 SUSY}. 
As discussed in \cite{Damgaard:2007be}, the orbifold projection 
can be achieved by restricting the fields corresponding to the $U(1)$
charges according to 
\begin{alignat}{2}
z_m &= \sum_{\fatk\in \Z_N^2}z_m(\fatk)\otimes E_{\fatk,\fatk+\fate_m}, 
&\quad 
\bar{z}_m &= \sum_{\fatk\in \Z_N^2}\bar{z}_m(\fatk)\otimes
 E_{\fatk+\fate_m,\fatk}, \nn \\
\label{restriction}
\eta  &= \sum_{\fatk\in \Z_N^2}\eta(\fatk)\otimes E_{\fatk,\fatk}, 
&\quad 
\psi_m &= \sum_{\fatk\in \Z_N^2}\psi_m(\fatk)\otimes
 E_{\fatk,\fatk+\fate_m}, \\
\chi_{12} &= \sum_{\fatk\in \Z_N^2}\chi_{12}(\fatk)\otimes
E_{\fatk+\fate_1+\fate_2,\fatk}, \nn 
\end{alignat}
where $E_{\bfk,\bfl}\equiv E_{k_1,\l_1}\otimes E_{k_2,l_2}$ 
with $\left(E_{ij}\right)_{kl}\equiv\delta_{ik}\delta_{jl}$.
As a result, we obtain the orbifolded action, 
\begin{align}
S_{\rm orb} = 
\frac{1}{g^2}{\rm Tr}\sum_{\mathbf k}\Biggl( 
&\frac{1}{4}\Bigl|z_m({\mathbf k})
z_n(\fatk+\fate_m) - z_n(\fatk)z_m(\fatk+\fate_n)\Bigr|^2  \cr
& + \frac{1}{8}\Bigl(z_m(\fatk)\bar{z}_m(\fatk)-\bar{z}_m(\fatk-\fate_m)z_m(\fatk
-\fate_m)\Bigr)^2 \cr
& + \eta(\fatk)
\Bigl(\bar{z}_m(\fatk-\bfe_m)\psi_m(\fatk-\bfe_m)
 -\psi_m(\fatk)\bar{z}_m(\fatk)\Bigr)\cr 
& - \frac{1}{2}\chi_{mn}(\fatk)
\Bigl(z_m(\fatk)\psi_n(\fatk+\bfe_m)
-\psi_n(\fatk)z_m(\fatk+\fate_n) \nn \\
&\hspace{2.3cm}-z_n(\fatk)\psi_m(\fatk+\fate_n)
+\psi_m(\fatk)z_n(\fatk+\fate_m)\Bigr).
\label{orbifold action}
\end{align}
A euclidean space-time lattice action for two-dimensional $\cN=(2,2)$
supersymmetric gauge theory is obtained by deconstruction: 
shifting $z_m(\bfk)$ and $\bz_m(\bfk)$ 
by $1/a$, where $a$ is a fundamental lattice spacing $a$ in the
orbifolded action  
(\ref{orbifold action})
\cite{Cohen:2003xe}\cite{Damgaard:2007be}. 
Another way of introducing the lattice spacing $a$ is to regard
the bosonic link variables $z_m(\bfk)$ and $\bz_m(\bfk)$ as
$\frac{1}{a}e^{iaA_m(\bfk)}$ and 
$\frac{1}{a}e^{-iaA_m^\dagger(\bfk)}$, respectively
\cite{Unsal:2006qp}, where $A_m(\bfk)$ are not hermitian but 
complex matrices. 
If we expand the action in $a$, the leading contribution clearly
agrees with the action which is obtained by ordinary deconstruction. 
In this procedure the action (\ref{orbifold action}) can be regarded 
as a lattice action for two-dimensional $\cN=(2,2)$
supersymmetric gauge theory. 

\begin{table}[h]
\caption{Two $U(1)$ charges}
\begin{center}
\begin{tabular}{c|cccccc}
 & $z_1$ & $z_2$ & $\eta$ & $\chi_{12}$ & $\psi_1$ & $\psi_2$ \\
\hline
${\mathbf r}$ & ${\mathbf e}_1$ & ${\mathbf e}_2$ & ${\mathbf 0}$ & 
-${\mathbf e}_1$-${\mathbf e}_2$ & ${\mathbf e}_1$ & ${\mathbf e}_2$ 
\end{tabular}
\end{center}
\label{remaining charge table 4 SUSY}
\end{table}

It is important to note that there exists
a trivial generalization of the lattice formulation 
(\ref{orbifold action}). 
In \cite{Cohen:2003xe} it is assumed that 
at least one fermion component has zero $U(1)$ charges in order to 
preserve at least one supersymmetry after orbifolding 
(see also \cite{Damgaard:2007be}). 
However, if we do not insist on preserving any supersymmetries, 
we can use the three independent $U(1)$ charges to obtain an orbifolded
action by linearly combining 
the $U(1)$ charges in Table \ref{charge table 4 SUSY}. 
The charge assignment in this case is summarized in Table 
\ref{three U(1) charges}.
\begin{table}[h]
\caption{Three $U(1)$ charges}
\begin{center}
\begin{tabular}{c|cccccc}
 & $z_1$ & $z_2$ & $\eta$ & $\chi_{12}$ & $\psi_1$ & $\psi_2$ \\
\hline
${\mathbf r}$ & ${\mathbf e}_1$ & ${\mathbf e}_2$ & $\bfa$ & 
$\bfa_{12}$ & ${\bfa}_1$ & ${\bfa}_2$ 
\end{tabular}
\end{center}
\label{three U(1) charges}
\end{table}
Here $\bfe_m$, $\bfa$, $\bfa_{12}$ and $\bfa_m$ are three-component
vectors with the relations, 
\begin{equation}
 \bfa+\bfa_m = \bfe_m, \quad \bfa_{12}+\bfa_m=-|\e_{mn}|\bfe_n, 
\quad \bfa+\bfa_1+\bfa_2+\bfa_{12}=0. 
\label{a-relations}
\end{equation}
Using this notation, we obtain the following more general orbifolded 
action:
\begin{align}
S_{\rm orb} = 
\frac{1}{g^2}{\rm Tr}\sum_{\mathbf k}\Biggl( 
&\frac{1}{4}\Bigl|z_m({\mathbf k})
z_n(\fatk+\fate_m) - z_n(\fatk)z_m(\fatk+\fate_n)\Bigr|^2  \nn \\
& +\frac{1}{8}\Bigl(z_m(\fatk)\bar{z}_m(\fatk)
 -\bar{z}_m(\fatk-\fate_m)z_m(\fatk-\fate_m)\Bigr)^2 \nn \\
& + \eta(\fatk)
\Bigl(\bar{z}_m(\bfk+\bfa-\bfe_m)\psi_m(\bfk+\bfa-\bfe_m)
-\psi_m(\fatk+\bfa)\bar{z}_m(\fatk+\bfa)\Bigr)\cr 
& - \frac{1}{2}\chi_{mn}(\fatk)
\Bigl(z_m(\fatk)\psi_n(\fatk+\fate_m)
-\psi_n(\fatk)z_m(\fatk+\bfa_n) \nn \\
&\hspace{2.3cm}-z_n(\fatk)\psi_m(\fatk+\fate_n)
+\psi_m(\fatk)z_n(\fatk+\bfa_m)\Bigr). 
\label{general orbifold action}
\end{align}
This is nothing but the action given in the link approach
\cite{D'Adda:2005zk} with the identifications%
\footnote{We understand that this equivalence was known to the 
authors of ref.\cite{D'Adda:2005zk}. 
(N.~Kawamoto, private communication)}, 
\begin{equation}
 z_m \equiv \sqrt{2} {\cal U}_{-m}, 
  \quad \bz_m \equiv \sqrt{2} {\cal U}_{+m}, \quad 
 \eta \equiv i\rho, \quad \chi_{12} \equiv i\tilde{\rho}, 
 \quad \psi_m \equiv \sqrt{2} \lambda_m, 
\label{identification}
\end{equation}
where the right hand sides correspond to the notation used in \cite{D'Adda:2005zk}. 
The relations (\ref{a-relations})
among $\bfe_m$, $\bfa$, $\bfa_m$ and $\bfa_{12}$
are also as given in \cite{D'Adda:2005zk}. We see that they are nothing 
but the charge assignments for the fields, and in particular
the shift variables $\bfa$, $\bfa_m$ and $\bfa_{12}$ are the
$U(1)$ charges of the fermions, a point of importance below.
A related issue pertains to the three-dimensional structure 
discussed in \cite{D'Adda:2004jb} and which can be understood 
in terms of the maximal 
number of $U(1)$ symmetries of the mother theory. 
In the following discussion, we concentrate for simplicity
on the case of $\bfa=0$.
It is straightforward to extend the discussion to the general case. 

We now turn to the question of preserved supersymmetries 
of the orbifolded theory. 
As discussed in \cite{Kaplan:2002wv}, the orbifolded action is
expected to be invariant only under the action of the
scalar supercharge $\hQ$, a singlet under all $U(1)$'s.
This is in agreement with the naive expectation that only
supersymmetries that do not generate space-time translations,
even discrete ones, can be preserved in general. 
One can see this explicitly as follows. Consider the
supersymmetry transformation 
(\ref{supertrans mother 2}) and the charge assignment of the fields. 
As we stressed above, the fermionic parameters 
$\hkappa$, $\hkappa_m$ and
$\hkappa_{12}$ must be proportional to the unit matrix
in order that (\ref{supertrans mother 2}) be a consistent
set of transformations that leave the action invariant.
After orbifolding this is simply impossible. If the corresponding
transformations in the orbifolded theory should be meaningful at
all, we are forced assign $U(1)$ charges ${\mathbf 0}$, $\bfe_m$ and 
$-\bfe_1-\bfe_2$ to $\hkappa$, $\hkappa_m$ and
$\hkappa_{12}$, respectively.
In order that the transformation (\ref{supertrans mother 2}) 
be consistent with 
the orbifold projection, the $\hkappa_A$ must thus take the form 
\begin{equation}
 \hkappa = \kappa {\mathbf 1}_{kN^2}, \quad 
 \hkappa_m = \kappa_m V_{\bfe_m}, \quad
 \hkappa_{12} = \kappa_{12} V_{-\bfe_1-\bfe_2},
 \label{kappa structure}
\end{equation}
where $\kappa$, $\kappa_m$ and $\kappa_{12}$ are Grassmann parameters 
and $V_\bfq$ is defined as
\begin{equation}
 V_{\bfq} \equiv \sum_{\bfk}{\mathbf 1}_{k}\otimes E_{\bfk,\bfk+\bfq}~.
\end{equation}
This is the essential reason why the supersymmetries corresponding 
to $\hQ_{12}$ and $\hQ_m$ are broken after orbifolding. 
In fact, as emphasized above, 
the ordinary variation $\delta S$ of the action (\ref{mother action 2}) 
under (\ref{supertrans mother 2}) is zero only when the supersymmetry
parameters are proportional to 
the unit matrix, using of course the usual Leibniz rule of variations,
\begin{equation}
 \delta (FG) = (\delta F)G + F (\delta G) ~. 
\label{usual Leibniz}
\end{equation}
This conventional Leibniz rule for the variations of matrices in the
mother theory leads to a modified rule 
for ${\hQ_A}=\{\hQ,\hQ_{12},\hQ_m\}$, 
\begin{equation}
 \hQ_A (FG) = (\hQ_A F)G + (-1)^{|F|} V_{A} F V_A^{-1} 
(\hQ_A G), 
\label{Leibniz for hat-Q}
\end{equation}
where $V_A$ expresses $1_{kN^2}$, $V_{-\bfe_1-\bfe_2}$ and $V_{\bfe_m}$ 
corresponding to $\hQ$, $\hQ_{12}$ and $\hQ_m$, respectively. 
It is easy to show that the action $S$ is not invariant under
the transformations generated by $\hQ_{12}$ and $\hQ_m$
due to the modified Leibniz rule (\ref{Leibniz for hat-Q}). 
Furthermore, with this modified rule the supersymmetry algebra 
(\ref{algebra}) is not satisfied when acting on the multiplet of
fields.
Therefore, only the supercharge $\hQ$ which is associated with $\hkappa$ 
is preserved after the orbifold projection. 

For the purpose of the discussion below,
let us define supercharges that act on lattice fields, the field variables
{\em after} orbifolding. 
Corresponding to the infinitesimal fermionic parameters (\ref{kappa structure}), 
we see that the supercharges 
can be expressed as matrices as well: 
\begin{equation}
 \hQ \equiv Q 1_{kN^2}, \quad 
 \hQ_{12} \equiv Q_{12} V_{\bfe_1+\bfe_2}, \quad
 \hQ_m \equiv Q_{m} V_{-\bfe_m}. 
\end{equation}
This definition arises from the fact that the variations 
(\ref{supertrans mother 2}) corresponding to the supersymmetry
transformations carry no $U(1)$ charges. 
The two expressions 
$\{\hQ_A\}=\{\hQ,\hQ_{12},\hQ_m\}$ and $\{Q_A\}=\{Q,Q_{12},Q_m\}$ 
are completely equivalent after orbifolding, but the former 
act on the large matrices $z_m, \cdots$ and the latter act on 
the lattice fields $z_m(\bfk),\cdots$. 
Using $\{\kappa_A\}$ and $\{Q_A\}$, the supersymmetry transformation 
can be combined into
\begin{equation}
 \delta \Phi = 2i \kappa Q\Phi -2i \kappa_{12} Q_{12}\Phi 
+2i \kappa_m Q_m\Phi ~. 
\label{SUSYQ}
\end{equation}
In terms of the lattice fields it 
can be written as 
\begin{align}
 \delta z_m(\bfk) = \,&2i\kappa \psi_m(\bfk) + 2i \kappa_m \eta(\bfk),
 \nn \\
 \delta \bz_m(\bfk) =\, &-2i \kappa_{mn}\psi_n(\bfk-\bfe_n)
  -2i \kappa_n \chi_{mn}(\bfk), \nn \\
 \delta \eta(\bfk) =\, &\frac{i}{2}\kappa 
  \Bigl(z_m(\bfk)\bz_m(\bfk)-\bz_m(\bfk-\bfe_m)z_m(\bfk-\bfe_m)\Bigr)
 \nn \\
 \label{supertrans orbifolded}
  &+i \kappa_{12}\Bigl(z_1(\bfk-\bfe_1-\bfe_2)z_2(\bfk-\bfe_2)
    -z_2(\bfk-\bfe_1-\bfe_2)z_1(\bfk-\bfe_1)\Bigr), \\
 \delta \chi_{12}(\bfk) =\, 
  &-i\kappa\Bigl(\bz_1(\bfk+\bfe_2)\bz_2(\bfk)
    -\bz_2(\bfk+\bfe_1)\bz_1(\bfk)\Bigr) \nn \\
  &-\frac{i}{2}\kappa_{12}
   \Bigl(z_m(\bfk)\bz_m(\bfk)-\bz_m(\bfk-\bfe_m)z_m(\bfk-\bfe_m)\Bigr),
 \nn \\
\delta \psi_m(\bfk) =\, &i\kappa_n\Bigl(
z_m(\bfk+\bfe_n)\bz_n(\bfk+\bfe_m)-\bz_n(\bfk)z_m(\bfk) \nn \\
&\hspace{1cm}-\frac{1}{2}\delta_{mn}
\Bigl(
z_l(\bfk)\bz_l(\bfk)-\bz_l(\bfk-\bfe_l)z_l(\bfk-\bfe_l)
\Bigr)
\Bigr). \nn
\end{align}
{}From the charge assignment for the supercharges, 
we see that $Q$, $Q_m$ and
$Q_{12}$ live on sites, links and diagonal links (or, equivalently,
corners), respectively. 
Therefore, the actions of $Q_m$ and $Q_{12}$ change the geometrical 
structure of operators. 
For example, $Q_m$ changes the link variable $z_m(\bfk)$ 
into a site variable $\eta(\bfk)$ as shown in the first line 
of (\ref{supertrans orbifolded}). 

We note that the operators $Q_A$ obey a usual Leibniz rule, 
\begin{equation}
 Q_A (F(\bfk)G(\bfk+\bfe_F))=\left(Q_A F(\bfk)\right)G(\bfk+\bfe_F)
+(-1)^{|F|}F(\bfk)\left(Q_A G(\bfk+\bfe_F)\right), 
\label{Leibniz for Q}
\end{equation}
where $F(\bfk)$ and $G(\bfk)$ are lattice fields obtained from 
matrices $F$ and $G$ with $U(1)$ charges $\bfe_F$ and $\bfe_G$, 
respectively. 
We stress that (\ref{Leibniz for Q}) is equivalent 
to (\ref{Leibniz for hat-Q}).

One notices that the there is 
an ambiguity in the definition of (\ref{supertrans orbifolded}). 
The transformation (\ref{supertrans orbifolded}) is determined 
from (\ref{supertrans mother 2}) using the definition 
(\ref{SUSYQ}). 
However, there is no a priori principle to determine the 
positions of $\hkappa_A$ in (\ref{supertrans mother 2}) 
and the rule of 
transformation (\ref{supertrans orbifolded}) depends on 
the positions of these fermionic parameters, 
since $\hkappa_A$ do not commute with other fields in general.

\section{Equivalence between the orbifolding procedure and the link approach}

In the previous section we have shown that the lattice action
given by the link approach can be completely reproduced 
by the orbifolding procedure. 
However, as explicitly demonstrated above, only the supercharge with 
zero $U(1)$ charges is preserved after orbifolding. The two
actions being identical, this presents
a puzzle in view of the arguments
\cite{D'Adda:2005zk}\cite{D'Adda:2007ax} 
that in the link approach 
{\em all} supersymmetries are preserved.
In this section, we show how also this claim can be understood 
in terms of the orbifolding procedure.

A first and interesting observation is that 
the transformations (\ref{supertrans orbifolded}) coincide
with those given of the link approach \cite{D'Adda:2005zk}
under the identification (\ref{identification}). 
Nevertheless,  
we cannot identify $Q_A$ with the supercharges in the link
approach, $s_A$.
The most important properties of $s_A$ are 
(1) they satisfy a modified Leibniz rule when acting on lattice
fields, 
\begin{equation}
 s_A (F(\bfk)G(\bfk+\bfe_F))=\left(s_A F(\bfk)\right)G(\bfk+\bfe_F)
+(-1)^{|F|}F(\bfk-\bfe_A)\left(s_A G(\bfk+\bfe_F)\right), 
\label{modified Leibniz in LA}
\end{equation}
and (2) they satisfy the supersymmetry algebra corresponding 
to (\ref{algebra}). 
However, the operators $Q_A$ do not possess both of these properties. 
In fact, the $Q_A$'s obey the usual Leibniz rule (\ref{Leibniz for Q}), 
and the only preserved part of the supersymmetry algebra is 
the one associated with nilpotency of the scalar charge $Q$,
as mentioned in the previous section.

However, the operators $Q_A$ turn out to satisfy the above 
two properties if we  
{\em impose} the usual Leibniz rule for $\hQ_A$. This is
potentially confusing, but it corresponds to imposing   
\begin{equation}
 \hQ_A(FG) = \left(\hQ_A F\right)G + (-1)^{|F|}F\left(\hQ_A G\right),
 \label{change Leibniz}
\end{equation}
instead of (\ref{Leibniz for hat-Q}), without altering the 
transformations (\ref{supertrans mother 2}). 
In fact, if we impose (\ref{change Leibniz}) by hand, 
we derive the correspondingly modified Leibniz rule for $Q_A$, 
\begin{equation}
 Q_A (F(\bfk)G(\bfk+\bfe_F))=\left(Q_A F(\bfk)\right)G(\bfk+\bfe_F)
+(-1)^{|F|}F(\bfk-\bfe_A)\left(Q_A G(\bfk+\bfe_F)\right), 
\label{modified rule in orb}
\end{equation}
which coincides with (\ref{modified Leibniz in LA}). 
Moreover, it is straightforward to see that $\hQ_A$ satisfy the 
supersymmetry algebra (\ref{algebra}) even after orbifolding 
if one imposes eq. (\ref{change Leibniz}). 
We conclude that the supercharges introduced in the 
link approach can be identified with the orbifolded supercharges 
of the mother theory (\ref{definition of Q}) after demanding by hand 
the unusual Leibniz rule (\ref{change Leibniz}). 
We note that this argument is unchanged under an assignment of 
non-zero $U(1)$ charges to $\eta$ as in (\ref{general orbifold action}). 
So the equivalence holds in general.

Although the supercharges $\hQ_A$ (or $Q_A$) 
with the unusual Leibniz rule 
do not generate supersymmetries in any usual sense,  
the modified Leibniz rule in the orbifolded theory 
(\ref{modified rule in orb}) is actually consistent with gauge
symmetry of the lattice theory. 
That is, the supersymmetry transformations (\ref{supertrans orbifolded}) 
commute with gauge transformations. 
As an example, let us consider a supersymmetry transformation, 
\begin{equation}
 Q_{12} \bz_1(\bfk) = \psi_2(\bfk-\bfe_2). 
\end{equation}
Since $\psi_2(\bfk)$ is a link variable, the gauge transformation 
of the right hand side is 
\begin{equation}
 \psi_2(\bfk-\bfe_2) \to g^{-1}(\bfk-\bfe_2)\psi_2(\bfk-\bfe_2)g(\bfk). 
\label{SUSY-gauge}
\end{equation}
On the other hand, let us first consider the gauge transformation 
of $\bz_1(\bfk)$, 
\begin{equation}
 \bz_1(\bfk) \to g^{-1}(\bfk+\bfe_1)\bz_1(\bfk)g(\bfk). 
\end{equation}
Recalling the modified rule (\ref{modified rule in orb}), 
we obtain 
\begin{equation}
 Q_{12}\bigl(g^{-1}(\bfk+\bfe_1)\bz_1(\bfk)g(\bfk)\bigr)
  = g^{-1}(\bfk-\bfe_2)\psi_2(\bfk-\bfe_2)g(\bfk), 
\end{equation}
which is the same as (\ref{SUSY-gauge}). 
This illustrates the fact that the action of $Q_A$ commutes 
with gauge transformation 
thanks to the modified Leibniz rule.


We close this section by pointing out that the question 
of possible additional 
symmetries of the orbifolded action appears even at the level 
of the mother theory, that is, in supersymmetric 
Yang-Mills matrix theory. 
As we have discussed, the supercharges of the link approach 
can be equivalently and compactly 
expressed as operators $\hQ_A$ that act on the large matrices in 
the orbifolded mother theory. 
{}From this point of view, all properties of the unusual supercharges 
$\hQ_A$ come from the matrix structure of the fermionic parameters 
$\hkappa_A$ as in eq. (\ref{kappa structure}) and the modified Leibniz rule 
for $\hQ_A$ as in eq. (\ref{change Leibniz}). 
An important observation is that we can consider the transformation 
(\ref{definition of Q}) with (\ref{kappa structure}) and 
(\ref{change Leibniz}) in the framework of the mother theory 
without reference to the orbifold projection.
Namely, 
we could imagine searching for additional symmetries of the mother
theory (or, one higher level up, in the $d$-dimensional theory for
which the mother theory is obtained by dimensional reduction) by
allowing the non-trivial fermionic $\kappa$-parameters 
(\ref{kappa structure}) and the modified Leibniz rule 
(\ref{change Leibniz}). 
The Leibniz rule of ordinary field variations 
is then also modified: 
\begin{equation}
\delta^{L}(FG) = \left(\delta^L F\right) G + 
V_L^{-1} F V_L \left(\delta^L G \right).
\end{equation}
The new examples based on
the link approach correspond, at the level of the mother theory
of matrices, precisely to this. This illustrates the problem
(or challenge)
in a quite transparent manner.

\section{Conclusions}

In this paper we have considered the relation between  
two lattice formulations of 
two-dimensional $\cN=(2,2)$ supersymmetric gauge theory: 
the orbifolding procedure given in \cite{Cohen:2003xe} and 
the link approach given in \cite{D'Adda:2005zk}. We have shown 
that the general action in the link approach can be obtained by 
the orbifolding procedure if one does not insist that one fermionic 
field has zero $U(1)$ charges. 
We have written down the would-be supersymmetry transformations after orbifolding,
and explicitly shown how they are broken by the projection. 
An interesting observation is that these transformations for  
the lattice fields coincide with those given in the link approach 
if one were allowed to introduce 
a matrix structure in the fermionic parameters
$\hkappa_A$. 
They do not correspond to symmetries of the action in any usual sense. 
We have also shown that, by imposing a modified Leibniz 
rule for the original supercharges by hand, 
the supercharges after orbifolding can be identified with 
those of the link approach. 
As a result, 
the formulations based on the link approach and the orbifolding 
are equivalent. Any symmetries of the former are also symmetries
of the latter, and vice versa.
We have pointed out that the same issue can be discussed in the framework of 
supersymmetric Yang-Mills matrix theory.



\vspace{0.5cm}
\noindent
{\sc Acknowledgement:}~ 
S.M. acknowledges support from a
JSPS Postdoctoral Fellowship for Research Abroad. The work of P.H.D. was
supported in parts by the European Community Network ENRAGE, grant number
MRTN-CT-2004-005616. We are grateful to N. Kawamoto and I. Kanamori for
discussions, and for their comments on a preliminary version of this
paper.

\bibliographystyle{JHEP}
\bibliography{refs}

\end{document}